\documentclass[a4paper,10pt]{article}

\usepackage[margin=25mm]{geometry}
\usepackage{times}
\usepackage{graphicx}
\usepackage{caption}
\usepackage{hyperref}
\usepackage{amsmath}
\usepackage{amssymb}
\usepackage{booktabs}
\usepackage{subcaption}
\usepackage{tikz}
\usetikzlibrary{shapes,arrows,positioning,calc,fit}
\usepackage{fancyhdr}
\usepackage{titlesec}
\titleformat{\section}{\bfseries\large\uppercase}{\thesection.}{1em}{}
\titleformat{\subsection}{\bfseries\normalsize}{\thesubsection}{1em}{}
\titleformat{\subsubsection}{\itshape\normalsize}{\thesubsubsection}{1em}{}

\renewenvironment{abstract}{
  \begin{center}
    {\large\bfseries Abstract \par}
  \end{center}
}{\par\vspace{2em}}

\usepackage[numbers]{natbib}

\usepackage{soul}

\begin{document}

\title{\bfseries Efficient prior sensitivity analysis for \\Bayesian model comparison}

\author{
Zixiao Hu$^{1, 2}$, Jason D. McEwen$^{1,2}$\\
{\small $^{1}$Mullard Space Science Laboratory (MSSL), University College London (UCL), Dorking RH5 6NT, UK}\\
{\small $^{2}$The Alan Turing Institute, London NW1 2DB, UK}
}

\date{}
\maketitle
\pagestyle{fancy}

\begin{abstract}
  Bayesian model comparison implements Occam's razor through its sensitivity to the prior. However, prior-dependence makes it important to assess the influence of plausible alternative priors. Such prior sensitivity analyses for the Bayesian evidence are expensive, either requiring repeated, costly model re-fits or specialised sampling schemes. By exploiting the learned harmonic mean estimator (LHME) for evidence calculation we decouple sampling and evidence calculation, allowing resampled posterior draws to be used directly to calculate the evidence without further likelihood evaluations.  This provides an alternative approach to prior sensitivity analysis for Bayesian model comparison that dramatically alleviates the computational cost and is agnostic to the method used to generate posterior samples.  We validate our method on toy problems and a cosmological case study, reproducing estimates obtained by full Markov chain Monte Carlo (MCMC) sampling and nested sampling re-fits. For the cosmological example considered our approach achieves up to $6000\times$ lower computational cost.
\end{abstract}

\section{Introduction}\label{sec:intro}
While much of modern statistical practice focuses on predictive performance, scientific discovery often necessitates selecting between discrete, theoretically motivated hypotheses \cite{breiman2001,shmueli2010}.
Explicit model comparison consequently persists as a cornerstone of the scientific method. Various approaches exist, including frequentist hypothesis testing, information criteria \cite{gelman2014}, and cross-validation \cite{piironen2017}, but Bayesian model comparison is popular in many fields due to its evaluation of \textit{relative plausibility} and principled \textit{penalisation of model complexity} \cite{kass1995,mackay2003,trotta2008}. Central to this endeavour is the calculation of the marginal likelihood, or \textit{Bayesian evidence}, which modern algorithms and software implementations have made feasible in many scenarios \cite{skilling2004,skilling2006,meng1996,gronau2020,mcewen2023,polanska2025}.

The Bayesian evidence (or just \textit{evidence}) embodies an automatic Occam's razor, naturally penalising complex models that do not sufficiently improve the fit to the data. The penalty arises from its definition as a prior predictive density. Models with vast, wasted parameter space, or those that place probability mass in physically forbidden regions, suffer a reduction in evidence, regardless of their maximum likelihood fit \cite{kass1995,mackay2003}.

However, this same mechanism imparts a direct sensitivity to the prior that warrants careful consideration. While the influence of the prior on parameter estimation typically ``washes out'' with increasing data, the evidence always remains sensitive to the prior volume \cite{sinharay2002}. Taken to the extreme, the evidence for any model can be made vanishingly small as the support of the prior goes to infinity (sometimes known as the Lindley-Bartlett paradox \cite{lindley1957,bartlett1957}). Bayesian evidences are often criticised for this fragility; if priors are arbitrary, the resulting model comparison is also arbitrary.

To mitigate this, best practice dictates that priors should be well-motivated \cite{gelman2017,llorente2023}. These include physical priors (e.g.\ non-negative mass or flux; \cite{carrillo2014purify}), uninformative Jeffreys priors that are invariant to a parameter transformation \citep{lee1989bayesian}, maximum entropy priors \cite{jaynes1968}, or informative priors for example to regularise inverse problems (e.g.\ \cite{price2021sparse}). Data-driven specifications include those potentially specified by a generative model (e.g.\ \cite{mcewen2023proximal,remy2023probabilistic, liaudat2024scalable, whitney2025}), and data-informed priors, where the posterior of an \textit{a priori} analysis is used as the prior for an analysis with new data (e.g.\ \cite{alsing2021nested}). Alternative formulations such as fractional \cite{ohagan1995} or intrinsic Bayes factors \cite{berger1996} also attempt to reduce prior dependency by performing model comparison with priors constructed from subsets of the data. However, as no prior specification is truly devoid of influence, it is important to explore the sensitivity of the evidence to reasonable alternative choices \cite{kass1995,gelman2013}.

Despite its importance, such sensitivity analysis is difficult to perform in practice. Formal approaches to sensitivity analysis for posterior inference exist, ranging from global methods that consider entire classes of priors \cite{berger1990} to local methods based on infinitesimal perturbations \cite{gustafson2000}, but these do not extend to Bayes factors. Conversely, the ``informal'' approach of checking discrete alternative priors is only tractable in simple limiting cases; one example is that under a highly informative likelihood, changing from one uniform prior to another leads the evidence to change by the ratio of the two constant prior densities (see Chapter 28.1 of \cite{mackay2003} and e.g.\ \cite{bevins2022,patel2024}). Some specific path-based sampling schemes for evidence calculation also allow sensitivity checks through re-weighting of a sequence of samples \cite{chopin2010,cameron2014}. Otherwise, in nearly all other cases prior dependence can only be assessed by expensive re-fitting. This latter case is the focus of this work.

To address this issue, we present an efficient approach to assess prior dependence by importance resampling existing posterior draws, which can be obtained from \textit{any} sampling method (e.g.\ any Markov chain Monte Carlo (MCMC) sampling strategy, neural density estimators, and others). Although this is a common technique in sensitivity analysis for posterior inference (\cite{berger1994,kallioinen2023}), its application to model comparison is underexplored since many effective techniques to compute the evidence are coupled to the sampling strategy, e.g.\ nested sampling \cite{skilling2004,skilling2006}. We leverage the learned harmonic mean estimator of the evidence \cite{mcewen2023,polanska2025} to decouple sampling and evidence calculation, enabling the resampled draws to be used directly for evidence calculation without further likelihood evaluations.

The remainder of the paper is structured as follows. Section 2 describes our method in detail, including diagnostics to flag unreliable importance sampling and identify when unnecessary recalculation can be avoided. We validate the method on toy problems and a cosmological case study in Section 3, demonstrating up to $6000\times$ speed-ups compared to full MCMC and nested sampling re-runs. Finally, concluding remarks are given in Section 4.

\section{Methods}\label{sec:methods}
We first review the learned harmonic mean estimator for evidence calculation (Section \ref{sec:harmonic}), then describe our resampling approach for prior sensitivity analysis (Section \ref{sec:sensitivity}).

\subsection{Bayesian evidence calculation with the learned harmonic mean estimator}\label{sec:harmonic}
Consider a model $M$ with parameters $\theta$, prior $p(\theta\mid M)$, and likelihood $L(D\mid\theta, M)$ given data $D$, for which we have obtained $N$ samples $\{\theta_i\}_{i=1}^N$ from the posterior $p(\theta\mid D, M)$ using any sampling method, such as MCMC or nested sampling.

The learned harmonic mean estimator (LHME) \cite{mcewen2023,polanska2025}, which builds on the approach of \cite{gelfand1994}, computes the Bayesian evidence by introducing a learned target distribution $\varphi(\theta)$ to stabilise the standard harmonic mean estimator \cite{newton1994,neal1994}:
\begin{equation}
  \hat{\rho} =
  \frac{1}{N} \sum_{i=1}^N \frac{\varphi(\theta_i)}{L(D \mid \theta_i, M)\,p(\theta_i\mid M)},
  \quad
  \theta_i \sim p(\theta \mid D, M).
  \label{eq:harmonic}
\end{equation}
where $\hat{\rho}$ is an unbiased estimator of the reciprocal evidence $\rho = 1/p(D\mid M)$ under the usual Monte Carlo expectation. If $\varphi(\theta)$ is the normalised posterior, $\hat{\rho}$ has zero variance and is therefore optimal \cite{mcewen2023}. Since this is intractable, the LHME approximates the optimal $\varphi(\theta)$ by learning a density estimator, such as a normalising flow trained on posterior samples \cite{papamakarios2021,polanska2025}. We emphasise that the learned $\varphi(\theta)$ need not be perfectly accurate, but critically must have thinner tails than the posterior for the variance of the estimator to be finite. Various strategies can be performed to ensure this \cite{mcewen2023,polanska2025}.  Recent approaches use normalising flows \cite{polanska2025}, where the variance of the base distribution of the normalising flow is reduced after training (lowering the ``temperature'' $T < 1$). This concentrates $\varphi(\theta)$ within the bulk of the posterior samples, avoiding the exploding variance problem of the original estimator.

\subsection{Prior sensitivity calculation}\label{sec:sensitivity}
We leverage the decoupling of sampling and evidence calculation that the LHME provides to reuse existing posterior samples for prior sensitivity analysis. 

Let us begin by describing the general procedure for reusing posterior draws for one model to approximate the evidence under an alternative model, before specialising to prior sensitivity. Consider an original model $M_0$ and an alternative model $M_1$, which are each defined by two likelihoods and priors that may or may not be the same. We are interested in the case where $M_0$ and $M_1$ are indexed by the same parameters $\theta$, but which have different distributional specifications. The posteriors for the models are:
\begin{equation}
  p(\theta\mid D, M_k) = \frac{L(D\mid\theta, M_k)\,p(\theta\mid M_k)}{p(D\mid M_k)},
  \quad k = 0, 1.
\end{equation}
Given samples $\{\theta_i\}_{i=1}^N$ from the original posterior $p(\theta \mid D, M_0)$, we can obtain samples from the alternative posterior $p(\theta \mid D, M_1)$ via \textit{sampling importance-resampling (SIR)} \cite{rubin1987,smith1992}, which proceeds as follows. Define the unnormalised importance weights as a ratio of the unnormalised posteriors:
\begin{equation}
  w_i = \frac{L(D\mid\theta_i, M_1)\,p(\theta_i\mid M_1)}{L(D\mid\theta_i, M_0)\,p(\theta_i\mid M_0)},
  \quad i = 1, \ldots, N.
\end{equation}
Furthermore, let 
\begin{equation}
  \tilde{w}_i = \frac{w_i}{\sum_{j=1}^N w_j}
\end{equation}
denote the self-normalised weights. We now obtain a new set of unweighted samples, which we will call $\{\theta^\ast_i\}_{i=1}^N$, by resampling from the original set ${\{\theta_i\}}_{i=1}^{N}$ with replacement, where the probability of selecting each $\theta_j$ is $\tilde{w}_j$. Equivalently, the resampled draws are i.i.d.\ from the empirical distribution
$\sum_{j=1}^N \tilde{w}_j\,\delta_{\theta_j}(\theta)$, where the $\delta_{\theta_j}(\theta)$ are delta functions centred at $\theta_j$. The samples $\theta^\ast_i$ are then distributed approximately according to the target distribution $p(\theta \mid D, M_1)$, since
\begin{equation}
  p(\theta \mid D, M_1) \propto \frac{L(D\mid\theta, M_1)\,p(\theta\mid M_1)}{L(D\mid\theta, M_0)~p(\theta\mid M_0)}\,p(\theta \mid D, M_0).
\end{equation}

For valid importance sampling, the support of the new posterior ($M_1$) must be contained within the original posterior ($M_0$). Specifically, the mass of the new posterior should be well-represented by the original samples; we discuss diagnostics to assess these conditions below.

Armed with the resampled draws, we can compute the evidence for the alternative model $M_1$ using the LHME. We define the target density $\varphi_1(\theta)$ for the new model $M_1$, which can be constructed either by reusing the original learned target $\varphi_0(\theta)$, or by re-training it on the resampled draws. The reciprocal evidence for $M_1$ can then be estimated as:
\begin{equation}
  \hat{\rho}_1 \approx \frac{1}{N}\sum_{i=1}^N \frac{\varphi_1(\theta_i^\ast)}{L(D\mid\theta_i^\ast, M_1)~p(\theta_i^\ast\mid M_1)},
  \quad
  \theta_i^\ast \sim \sum_{j=1}^N \tilde{w}_j~\delta_{\theta_j}(\theta).
  \label{eq:reweighted_evidence}
\end{equation}

We now specialise to the case of prior sensitivity analysis, where the models $M_0$ and $M_1$ share the same likelihood function but differ in their prior specifications. The importance weights simplify to:
\begin{equation}
  w_i = \frac{p(\theta_i\mid M_1)}{p(\theta_i\mid M_0)},
  \quad i = 1, \ldots, N
\end{equation}
since the likelihood terms cancel. The resampled draws $\{\theta_i^\ast\}$ are then approximately distributed according to the posterior under the new prior $p(\theta \mid D, M_1)$. 

The new target $\varphi_1(\theta)$ is then constructed as follows. If the posterior is relatively insensitive to the prior change (guidance for determining this is given below), we reuse the original learned target as the new target, so that $\varphi_1(\theta) = \varphi_0(\theta)$. The evidence for the alternative prior is then computed immediately with Eq.~\ref{eq:reweighted_evidence}. This is extremely cheap, requiring only evaluations of the new prior on existing samples. On the other hand, if the posterior changes significantly we re-train $\varphi_1(\theta)$ on the resampled draws. This can be fine-tuned from the original learned target for faster convergence, and typically takes only tens of seconds to a few minutes for low-dimensional problems. In computationally demanding settings, this is still orders of magnitude faster than a full model re-fit. For clarity, (model) \textit{re-fit} refers to re-running the entire sampling procedure to obtain new posterior samples, while \textit{re-train} refers to only re-training the normalising flow target $\varphi(\theta)$ on resampled draws.

We employ several diagnostics to assess the reliability of the estimates.
\begin{itemize}
  \item First, we monitor the Pareto-$\hat{k}$ diagnostic, which estimates the shape parameter $k$ of the generalised Pareto distribution fitted to the upper tail of the importance weights \cite{vehtari2024}. Importance sampling is typically empirically unreliable if $\hat{k}$ exceeds $0.7$, so we adopt this as a heuristic to indicate that model-refit is necessary.
  \item Second, we use the effective sample size (ESS) of the importance weights,
        \begin{equation}
          \text{ESS} = \frac{\left(\sum_{i=1}^N w_i\right)^2}{\sum_{i=1}^N w_i^2} ,
        \end{equation}
        to determine whether re-training of $\varphi(\theta)$ is required. Since the fractional ESS measures the overlap between the original and new posteriors, we recommend re-training if this drops below $0.95$ (as a conservative default) to maintain low variance in the evidence estimate.
\end{itemize}
These checks are performed in addition to standard LHME diagnostics (see \cite{mcewen2023} for details). We also find that if the posterior is insensitive to the prior and weights are all equal, Pareto fitting diverges. In this case, when the maximum and minimum weights are equal to within some tolerance, we return $\hat{k}=-\infty$ directly as the $\hat{k}$ that reproduces a delta distribution for the weights in the limit.

We note that the above approach can be straightforwardly applied to other stabilised harmonic mean estimators (e.g.\ \cite{robert2009,metodiev2025,naderi2025}), which differ in the construction of the target $\varphi(\theta)$, as well as other methods that require only posterior samples to compute the evidence \cite{heavens2017,jia2020,srinivasan2024,rinaldi2024}. This work focuses on the learned variant due to its well-tested performance in real applications \cite{spuriomancini2023,piras2024,polanska2024,spuriomancini2024,paradiso2024,stiskalek2024,carrion2025,du2025,stiskalek2025}.

\begin{figure*}[t]
  \centering
  \begin{tikzpicture}[
      inputs/.style={rectangle, draw, fill=yellow!10, text centered, minimum height=3em, rounded corners, inner sep=5pt},
      block/.style={rectangle, draw, fill=blue!10, text centered, minimum height=3em, inner sep=3pt},
      decision/.style={diamond, draw, fill=green!10, text width=6em, text centered, inner sep=3pt, aspect=2},
      line/.style={draw, -latex', thick, scale=8}
    ]

    \node [inputs, text width=10.0em] (samples) {Original samples $\{\theta_i\} \sim p(\theta\mid D, M_0)$ with prior $p(\theta\mid M_0)$};
    \node [inputs, right=of samples, node distance=0.25cm] (prior) {\shortstack{Alternative prior\\$p(\theta\mid M_1)$}};

    \node [block, below=of samples, node distance=1.5cm, text width=8.0em, inner sep=3pt] (reweight) {Compute weights $w_i = \frac{p(\theta_i\mid M_1)}{p(\theta_i\mid M_0)}$ and resample};

    \node [decision, below=of reweight, node distance=1.5cm] (check_k) {$\hat{k} < 0.7$?};

    \node [block, left=of check_k, fill=red!20, text width=6.5em, xshift=-0.5cm] (refit) {Full model re-fit required};

    \node [decision, below=of check_k] (check_ess) {Fractional ESS $> 0.95$?};

    \node [block, right=of check_ess, text width=6em, node distance=0.5cm] (re-train) {Re-train target $\varphi(\theta)$ };

    \node [block, below=of check_ess, node distance=1.5cm] (compute) {Compute evidence (Eq.~\ref{eq:reweighted_evidence})};
    \node [inputs, below=of compute, node distance=1cm, minimum height=2.5em] (output) {Reciprocal evidence for alternative prior $\hat{\rho}_1$};

    \path [line] (samples) -- (reweight);
    \path [line] (prior) |- (reweight);
    \path [line] (reweight) -- (check_k);

    \path [line] (check_k) -- node [above, xshift=0.5cm, font=\small] {No} (refit);
    \path [line] (check_k) -- node [right, font=\small] {Yes} (check_ess);

    \path [line] (check_ess) -- node [above, font=\small] {No} (re-train);
    \path [line] (check_ess) -- node [right, font=\small] {Yes} (compute);

    \path [line] (re-train) |- (compute);

    \path [line] (compute) -- (output);
    \node [draw, dashed, rectangle, thick, inner sep=0.5cm, fit=(reweight)(check_k)(check_ess)(re-train)  (compute)] (algobox) {};
  \end{tikzpicture}
  \caption{Schematic of the prior sensitivity workflow. We first resample existing posterior draws to the posterior under the new prior. We then perform a two-stage diagnostic check: 1.\ if the Pareto-$\hat{k}$ diagnostic exceeds 0.7, the importance weights are unreliable and a full model re-fit is required; 2.\ if $\hat{k}$ is safe but fractional ESS is low ($<0.95$), we re-train the learned target $\varphi(\theta)$ to maintain low variance (blue). After the two checks, the evidence for the alternative prior is computed by Eq.~\ref{eq:reweighted_evidence}.}
  \label{fig:method_flowchart}
\end{figure*}
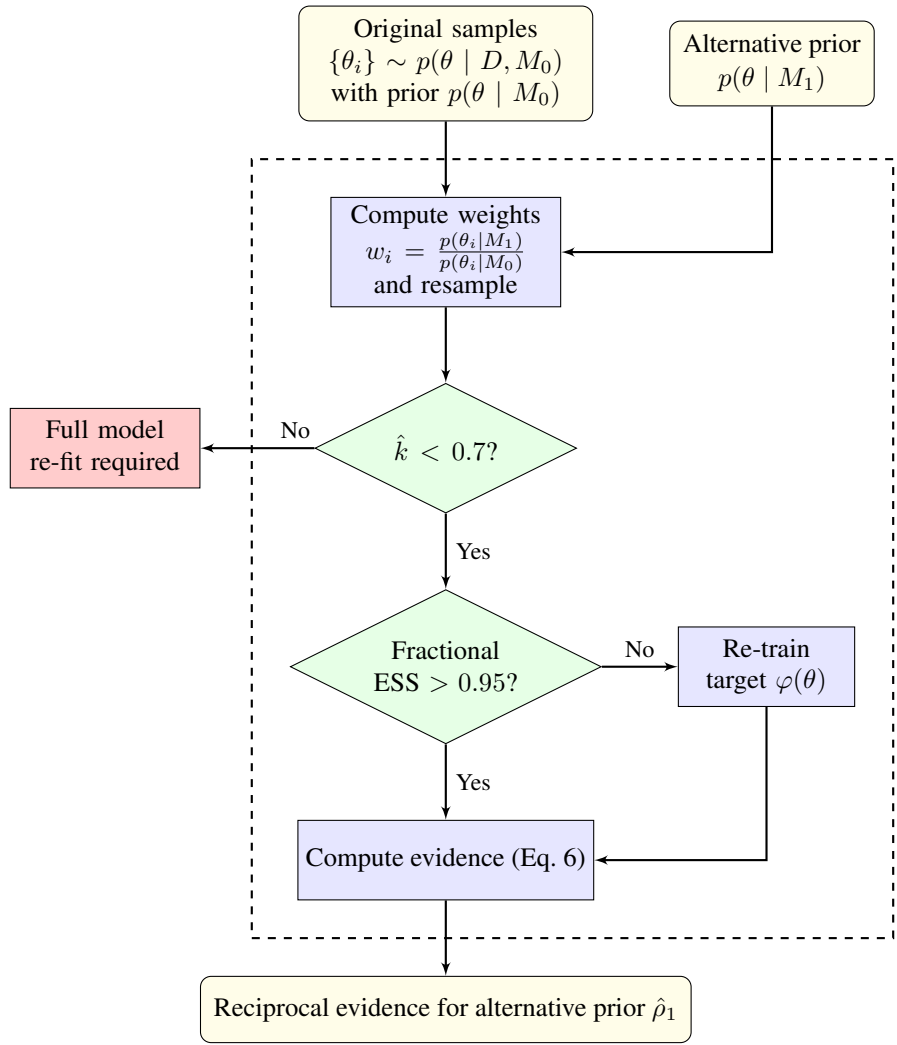

A schematic of the overall workflow is shown in Fig. \ref{fig:method_flowchart}.

\section{Results}
We first present toy experiments to validate accuracy and diagnostics, then apply the method to a cosmological case study, comparing against full MCMC and nested sampling re-fits and report computational savings. In what follows, we report the natural logarithm of the evidence, $\log Z = \log p(D\mid M)$, whose estimate and uncertainty are obtained by inverting $\hat{\rho}$ and performing standard error propagation \cite{mcewen2023,polanska2025}.

\subsection{Toy examples}
We demonstrate the method on two toy examples where the ground truth evidence under alternative priors is available for comparison, and assess the performance of the diagnostics.

\begin{figure}
  \subfloat[Gaussian densities]{
    \includegraphics[width=0.53\textwidth]{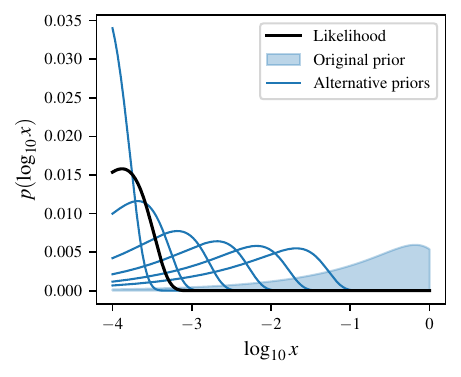}
    \label{fig:gaussian_densities}
  }
  \hfill
  \subfloat[Rosenbrock contours]{
    \includegraphics[width=0.43\textwidth]{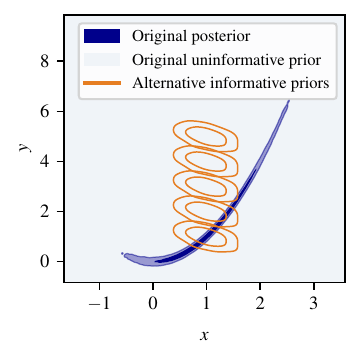}
    \label{fig:rosenbrock_contours}
  }
  \caption{(a) Illustration of prior widths versus the likelihood for the Gaussian toy example, in terms of the 1D marginal density (log-scale). (b) Rosenbrock posterior contours under an uninformative uniform prior (navy) and contours for alternative informative Gaussian priors (red) for different shifts along the $y$-axis. The original uninformative prior (light blue) covers the entire plotted region.}
\end{figure}

\textbf{Gaussian in 10 dimensions.} We first consider a 10-dimensional Gaussian problem to assess sensitivity to changes in prior strength. We use a Gaussian likelihood with zero mean and diagonal covariance (standard deviation $\sigma = 2 \times 10^{-4}$) and an original prior with zero mean and diagonal covariance ($\sigma = 1.0$). Posterior samples are taken with the No-U-Turn Sampler (NUTS) \cite{hoffman2014} for 16 chains of 1000 samples. We then evaluate the evidence under a set of alternative zero-mean Gaussian priors with widths ranging from $\sigma = 10^{-1.5}$ to $ 10^{-4}$, as shown in Fig. \ref{fig:gaussian_densities}. The analytical evidence is available in closed form for comparison.

\begin{figure*}
  \centering
  \includegraphics[width=\textwidth]{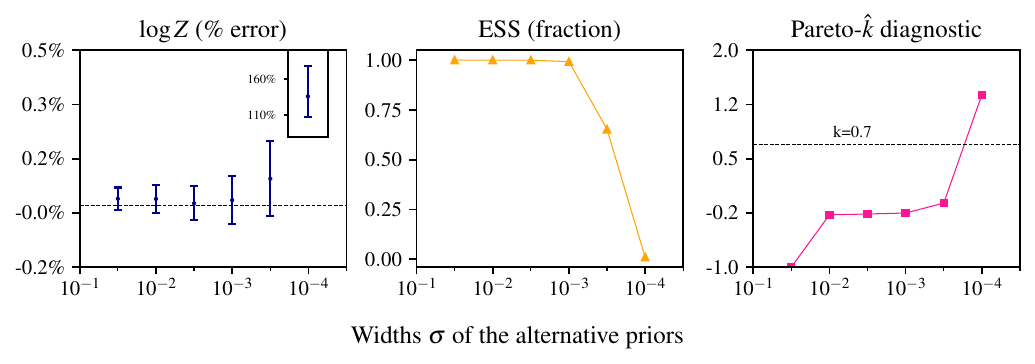}
  \caption{Prior dependence for the Gaussian example. The width of the likelihood is $2\times 10^{-4}$, which is between the narrowest and second narrowest priors considered. Left: Percent errors in the evidence computed with the LHME using importance resampled posterior draws, compared to the analytical solution as the strength of the prior is varied. Middle: Fractional ESS of the importance sampling. Right: Pareto-$\hat{k}$ diagnostic values for importance sampling. The evidence estimate is correctly flagged as unreliable when $\hat{k}$ is high (above 0.7).}
  \label{fig:gaussian_results}
\end{figure*}

Fig. \ref{fig:gaussian_results} shows the resampled evidence estimates and diagnostics. We report $\log Z$, the (non-reciprocal) log-evidence. The posterior is insensitive to the prior change for the four widest alternative priors, maintaining a fractional ESS close to 1.0.  No re-training of $\varphi(\theta)$ is necessary. The second narrowest prior approaches the scale of the likelihood, so fractional ESS drops to around 0.65 and re-training of $\varphi(\theta)$ is required. Finally, the narrowest prior is more informative than the likelihood, causing the posterior to drastically change (ESS $\approx 0$). Here, the importance sampled evidence estimate deviates significantly from the analytical value, which is correctly diagnosed by a Pareto-$\hat{k} > 0.7$.

\textbf{Rosenbrock in two dimensions.} We now assess the evidence for a sequence of informative priors against a broad baseline using the standard Rosenbrock likelihood ($a=1$, $b=100$). The baseline prior is uniform on $[-10,10]^2$, and the original posterior was obtained with NUTS (32 chains of 2000 samples). For the alternative hypotheses we use narrow Gaussian priors ($\Sigma\approx0.06$): the first is centred at the likelihood mode $(1,1)$ and subsequent priors are shifted along the $y$-axis to emulate priors that increasingly conflict with the data. Contours of the posterior and alternative priors are shown in Fig.~\ref{fig:rosenbrock_contours}.

\begin{figure*}
  \centering
  \includegraphics[width=\textwidth]{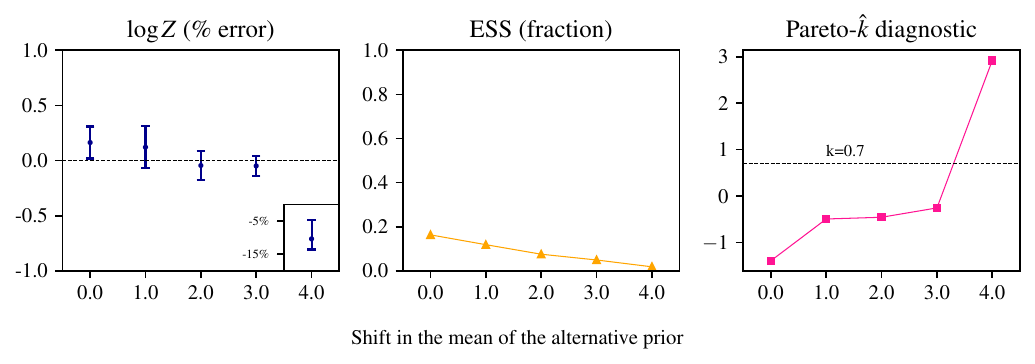}
  \caption{Same as Fig. \ref{fig:gaussian_results}, but for the Rosenbrock example with informative priors at different locations. Here ESS is low throughout and re-training $\varphi(\theta)$ is necessary. Pareto-$\hat{k} > 0.7$ successfully identifies the case where the evidence estimate is unreliable.}
  \label{fig:rosenbrock_results}
\end{figure*}

Evidence estimates with diagnostics appear in Fig.~\ref{fig:rosenbrock_results}. Here, ESS is low and re-training of $\varphi(\theta)$ is necessary for all priors. All evidences are consistent with ground truth (obtained by Riemann integration across a fixed grid) except for the prior with maximal shift, which is correctly flagged by Pareto-$\hat{k}$ as unreliable.


\subsection{Cosmological example}
We apply our method to a cosmological case study comparing logarithmic and uniform priors for a scale parameter, which has invoked significant debate in the literature \cite{simpson2017,schwetz2017,hergt2021,jimenez2022,gariazzo2022}. We consider the case of the tensor-to-scalar ratio of primordial perturbations $r$, reusing (Metropolis-Hastings) posterior samples provided by \cite{hergt2021} and comparing against their reported nested sampling evidences.

The original analysis in \cite{hergt2021} considers three different priors for $r$:
\begin{equation}
  \log_{10} r \sim \mathcal{U}(-10, 0), \quad
  \log_{10} r \sim \mathcal{U}(-5, 0), \quad \text{and} \quad
  r \sim \mathcal{U}(0, 1).
\end{equation}
We start with MCMC samples for the $\log_{10} r \sim \mathcal{U}(-10, 0)$ prior as the original posterior, and importance resample to the two alternative priors to obtain the evidences. We note that the error estimate of the LHME relies on the variance of the mean across from many MCMC chains, which can be noisy when the number of chains is low. A more reliable uncertainty can be obtained by bootstrap resampling the chains, taking at least 30 replicate chains, and computing the resulting standard deviation. The samples provided by \cite{hergt2021} have 8 chains, so we adopt this approach. While this takes longer than the standard error estimate, computational costs remains on the order of minutes and orders of magnitude faster than a full model re-fit.

We recover evidences consistent with standard error to both the nested sampling re-fits reported in \cite{hergt2021}, and MCMC re-fits combined with LHME, as shown in Table~\ref{tab:cosmo_evidence}. The variance is higher for the $r \sim \mathcal{U}(0, 1)$ prior due to the larger change from the original posterior, but remains consistent and sufficient for practical use where order unity changes in log evidence are of interest.
For the two alternative priors, the fractional ESS are $0.41$ and $0.063$ respectively, indicating moderate and significant changes in the posterior, so we re-train $\varphi(\theta)$ in both cases. Pareto-$\hat{k}$ diagnostics are $-\infty$ and $0.10\pm0.01$, both below the $0.7$ threshold. $\hat{k} = -\infty$ for the $\log_{10} r \sim \mathcal{U}(-5, 0)$ alternative prior indicates that the posterior is insensitive to the prior change, so the reduced ESS is due to the change in boundary of the prior support rather than a significant shift in the posterior mass.

The computational savings for our method are listed in the rightmost column of Table \ref{tab:cosmo_evidence}. The original MCMC chains were run on 32 CPU cores, so we report CPU core-hours for consistency. Timings for the nested sampling runs were not recorded, but are typically within similar order of magnitude to MCMC. The resampled evidence estimates with bootstrapped uncertainties take 10 and 30 minutes on a single CPU core for the two alternative priors respectively, which is around $3000\times$ and $6000\times$ faster than the full re-fits.

A minor note is that the nested sampling evidences we report here differ slightly from \cite{hergt2021}, because cosmological theory codes typically reject certain unphysical parameter combinations at the level of the prior. The raw nested sampling output is instead the evidence for a prior restricted to the physically allowed region. Quantities shown in Table \ref{tab:cosmo_evidence} are the corrected values for the actual prior used, which were reported in the code release of \cite{hergt2021} but not the paper. The difference is not qualitatively significant, and one could even argue that the uncorrected evidence is more meaningful as a prior only over the physically allowed parameter space (following the discussion in Section \ref{sec:intro}). Nevertheless, we report the corrected version here for consistency between methods.

\begin{table}[h]
  \centering
  \caption{Log-evidence estimates ($\log Z$) for the cosmological model under different priors. Our resampling method (bottom row in each block) reproduces evidence calculations to within the quoted uncertainties, while avoiding computationally expensive full re-runs. Uncertainties for the resampled estimates are obtained via bootstrap resampling of the original MCMC samples. The rightmost column shows computational time in CPU core-hours for each method.  Our resampled LHME approach provides savings in computational cost of around $3000\times$ and $6000\times$.}
  \label{tab:cosmo_evidence}
  \begin{tabular}{@{} l l r r @{} }
    \toprule
    \textbf{Prior}                        & \textbf{Method}         & \multicolumn{1}{c}{$\boldsymbol{\log Z}$} & \textbf{Time (CPU-hours)} \\
    \midrule
    Original                              & Nested sampling         & $-1432.43 \pm 0.22$                       & --                        \\
    $\log_{10} r \sim \mathcal{U}(-10,0)$ & MCMC + LHME             & $-1432.75 \pm 0.01$                       & 1152                      \\
    \midrule
    Alternative 1                         & Nested sampling         & $-1432.98 \pm 0.22$                       & --                        \\
    $\log_{10} r \sim \mathcal{U}(-5, 0)$ & MCMC + LHME             & $-1432.98 \pm 0.02$                       & 1440                      \\
                                          & {Resampled LHME (ours)} & ${-1432.94 \pm 0.06}$                     & {0.5}                     \\
    \midrule
    Alternative 2                         & Nested sampling         & $-1435.44 \pm 0.23$                       & --                        \\
    $r \sim \mathcal{U}(0, 1)$            & MCMC + LHME             & $-1435.57 \pm 0.02$                       & 1056                      \\
                                          & {Resampled LHME (ours)} & ${-1435.43 \pm 0.33}$                     & {0.16}                    \\
    \bottomrule
  \end{tabular}
\end{table}

\section{Conclusion}
We present a fast method to assess prior sensitivity of the Bayesian evidence, which is agnostic to the method used to generate samples, by resampling existing posterior draws. A two-stage diagnostic check for unreliable importance sampling is introduced to avoid unnecessary recalculation and check for failure. We validate on toy problems and a cosmological case study, reproducing Bayesian evidences obtained from full MCMC or nested sampling runs at several thousand times lower computational cost.

The learned harmonic mean estimator makes this efficient post-hoc approach possible by decoupling sampling and evidence calculation. Other post-hoc computational schemes that make use of these advantages are in development, such as to combine evidences across datasets, requiring only posterior samples from each dataset obtained in any manner \cite{hu2026}.

Code for the method will be made publicly available in the \texttt{harmonic}~\footnote{\url{https://github.com/astro-informatics/harmonic}} package. We hope that the increased accessibility brought by the method will encourage transparent prior dependence reporting in Bayesian model comparison applications, which is critical for robust scientific inferences.

\bibliographystyle{unsrt}
\bibliography{references}

@article{bartlett1957,
  title   = {A Comment on {{D}}. {{V}}. {{Lindley}}'s Statistical Paradox},
  author  = {Bartlett, M. S.},
  year    = 1957,
  month   = dec,
  journal = {Biometrika},
  volume  = {44},
  number  = {3-4},
  pages   = {533--534},
  issn    = {0006-3444},
  doi     = {10.1093/biomet/44.3-4.533},
  urldate = {2025-11-19}
}

@article{lindley1957,
  title   = {A {{Statistical Paradox}}},
  author  = {Lindley, D. V.},
  year    = 1957,
  month   = jun,
  journal = {Biometrika},
  volume  = {44},
  number  = {1-2},
  pages   = {187--192},
  issn    = {0006-3444},
  doi     = {10.1093/biomet/44.1-2.187},
  urldate = {2025-11-19}
}

@misc{mcewen2023,
  title         = {Machine Learning Assisted {{Bayesian}} Model Comparison: Learnt Harmonic Mean Estimator},
  shorttitle    = {Machine Learning Assisted {{Bayesian}} Model Comparison},
  author        = {McEwen, Jason D. and Wallis, Christopher G. R. and Price, Matthew A. and Mancini, Alessio Spurio},
  year          = 2023,
  month         = nov,
  number        = {arXiv:2111.12720},
  eprint        = {2111.12720},
  publisher     = {arXiv},
  doi           = {10.48550/arXiv.2111.12720},
  urldate       = {2024-10-09},
  archiveprefix = {arXiv}
}

@article{polanska2025,
  title     = {Learned Harmonic Mean Estimation of the {{Bayesian}} Evidence with Normalizing Flows},
  author    = {Polanska, Alicja and Price, Matthew A. and Piras, Davide and Mancini, Alessio Spurio and McEwen, Jason D.},
  year      = 2025,
  month     = oct,
  journal   = {The Open Journal of Astrophysics},
  volume    = {8},
  publisher = {Maynooth Academic Publishing},
  doi       = {10.33232/001c.146026},
  urldate   = {2025-11-10},
  langid    = {english}
}

@article{vehtari2024,
  title   = {Pareto {{Smoothed Importance Sampling}}},
  author  = {Vehtari, Aki and Simpson, Daniel and Gelman, Andrew and Yao, Yuling and Gabry, Jonah},
  year    = 2024,
  journal = {Journal of Machine Learning Research},
  volume  = {25},
  number  = {72},
  pages   = {1--58},
  issn    = {1533-7928},
  urldate = {2025-11-19}
}

@article{kass1995,
  title      = {Bayes {{Factors}}},
  shorttitle = {Prior Sensitivity},
  author     = {Kass, Robert E. and Raftery, Adrian E.},
  year       = 1995,
  journal    = {J. Am. Statist. Assoc.},
  volume     = {90},
  number     = {430},
  pages      = {773--795},
  doi        = {10.1080/01621459.1995.10476572}
}

@article{chopin2010,
  title   = {Properties of Nested Sampling},
  author  = {Chopin, Nicolas and Robert, Christian P.},
  year    = 2010,
  month   = sep,
  journal = {Biometrika},
  volume  = {97},
  number  = {3},
  pages   = {741--755},
  issn    = {0006-3444},
  doi     = {10.1093/biomet/asq021},
  urldate = {2025-11-19}
}

@article{cameron2014,
  title     = {Recursive {{Pathways}} to {{Marginal Likelihood Estimation}} with {{Prior-Sensitivity Analysis}}},
  author    = {Cameron, Ewan and Pettitt, Anthony},
  year      = 2014,
  month     = aug,
  journal   = {Statistical Science},
  volume    = {29},
  number    = {3},
  pages     = {397--419},
  publisher = {Institute of Mathematical Statistics},
  issn      = {0883-4237, 2168-8745},
  doi       = {10.1214/13-STS465},
  urldate   = {2025-11-19}
}

@book{gelman2013,
  title     = {Bayesian {{Data Analysis}}},
  author    = {Gelman, Andrew and Carlin, John B. and Stern, Hal S. and Dunson, David B. and Vehtari, Aki and Rubin, Donald B.},
  year      = 2013,
  month     = nov,
  edition   = {3},
  publisher = {{Chapman and Hall/CRC}},
  address   = {New York},
  doi       = {10.1201/b16018},
  isbn      = {978-0-429-11307-9}
}

@book{mackay2003,
  title       = {Information {{Theory}}, {{Inference}} and {{Learning Algorithms}}},
  author      = {MacKay, David J. C.},
  year        = 2003,
  month       = sep,
  publisher   = {Cambridge University Press},
  googlebooks = {AKuMj4PN\_EMC},
  isbn        = {978-0-521-64298-9},
  langid      = {english}
}

@article{breiman2001,
  title      = {Statistical {{Modeling}}: {{The Two Cultures}} (with Comments and a Rejoinder by the Author)},
  shorttitle = {Statistical {{Modeling}}},
  author     = {Breiman, Leo},
  year       = 2001,
  month      = aug,
  journal    = {Statistical Science},
  volume     = {16},
  number     = {3},
  pages      = {199--231},
  publisher  = {Institute of Mathematical Statistics},
  issn       = {0883-4237, 2168-8745},
  doi        = {10.1214/ss/1009213726},
  urldate    = {2025-11-19}
}

@article{shmueli2010,
  title     = {To {{Explain}} or to {{Predict}}?},
  author    = {Shmueli, Galit},
  year      = 2010,
  month     = aug,
  journal   = {Statistical Science},
  volume    = {25},
  number    = {3},
  pages     = {289--310},
  publisher = {Institute of Mathematical Statistics},
  issn      = {0883-4237, 2168-8745},
  doi       = {10.1214/10-STS330},
  urldate   = {2025-11-19}
}

@article{berger1990,
  title      = {Robust {{Bayesian}} Analysis: Sensitivity to the Prior},
  shorttitle = {Robust {{Bayesian}} Analysis},
  author     = {Berger, James O.},
  year       = 1990,
  month      = jul,
  journal    = {Journal of Statistical Planning and Inference},
  volume     = {25},
  number     = {3},
  pages      = {303--328},
  issn       = {0378-3758},
  doi        = {10.1016/0378-3758(90)90079-A},
  urldate    = {2025-11-19}
}

@article{trotta2008,
  title         = {Bayes in the Sky: {{Bayesian}} Inference and Model Selection in Cosmology},
  shorttitle    = {Bayes in the Sky},
  author        = {Trotta, Roberto},
  year          = 2008,
  month         = mar,
  journal       = {Contemporary Physics},
  volume        = {49},
  number        = {2},
  eprint        = {0803.4089},
  primaryclass  = {astro-ph},
  pages         = {71--104},
  issn          = {0010-7514, 1366-5812},
  doi           = {10.1080/00107510802066753},
  urldate       = {2025-05-15},
  archiveprefix = {arXiv},
  langid        = {english}
}

@article{skilling2006,
  title     = {Nested Sampling for General {{Bayesian}} Computation},
  author    = {Skilling, John},
  year      = 2006,
  month     = dec,
  journal   = {Bayesian Analysis},
  volume    = {1},
  number    = {4},
  pages     = {833--859},
  publisher = {International Society for Bayesian Analysis},
  issn      = {1936-0975, 1931-6690},
  doi       = {10.1214/06-BA127},
  urldate   = {2025-11-10}
}

@article{newton1994,
  title      = {Approximate {{Bayesian Inference}} with the {{Weighted Likelihood Bootstrap}}},
  author     = {Newton, Michael A. and Raftery, Adrian E.},
  year       = 1994,
  journal    = {Journal of the Royal Statistical Society. Series B (Methodological)},
  volume     = {56},
  number     = {1},
  eprint     = {2346025},
  eprinttype = {jstor},
  pages      = {3--48},
  publisher  = {[Royal Statistical Society, Oxford University Press]},
  issn       = {0035-9246},
  urldate    = {2025-11-20}
}

@article{gelman2017,
  title     = {The {{Prior Can Often Only Be Understood}} in the {{Context}} of the {{Likelihood}}},
  author    = {Gelman, Andrew and Simpson, Daniel and Betancourt, Michael},
  year      = 2017,
  month     = oct,
  journal   = {Entropy},
  volume    = {19},
  number    = {10},
  pages     = {555},
  publisher = {Multidisciplinary Digital Publishing Institute},
  issn      = {1099-4300},
  doi       = {10.3390/e19100555},
  urldate   = {2025-11-20},
  copyright = {http://creativecommons.org/licenses/by/3.0/},
  langid    = {english}
}

@article{hergt2021,
  title         = {Bayesian Evidence for the Tensor-to-Scalar Ratio {$r$} and Neutrino Masses {$m_\nu$}: Effects of Uniform vs Logarithmic Priors},
  shorttitle    = {Bayesian Evidence for the Tensor-to-Scalar Ratio {$r$} and Neutrino Masses {$m_\nu$}},
  author        = {Hergt, Lukas T. and Handley, Will J. and Hobson, Michael P. and Lasenby, Anthony N.},
  year          = 2021,
  month         = jun,
  journal       = {Physical Review D},
  volume        = {103},
  number        = {12},
  eprint        = {2102.11511},
  primaryclass  = {astro-ph},
  pages         = {123511},
  issn          = {2470-0010, 2470-0029},
  doi           = {10.1103/PhysRevD.103.123511},
  urldate       = {2025-10-01},
  archiveprefix = {arXiv}
}

@article{neal1994,
  author  = {Neal, Radford M.},
  journal = {JR Stat Soc Ser A (Methodological)},
  pages   = {41--42},
  title   = {{Contribution to the discussion of “Approximate Bayesian inference with the weighted likelihood bootstrap” by Newton MA, Raftery AE}},
  volume  = {56},
  year    = {1994}
}

@article{hoffman2014,
  title      = {The {{No-U-turn}} Sampler: Adaptively Setting Path Lengths in {{Hamiltonian Monte Carlo}}},
  shorttitle = {The {{No-U-turn}} Sampler},
  author     = {Hoffman, Matthew D. and Gelman, Andrew},
  year       = 2014,
  month      = jan,
  journal    = {J. Mach. Learn. Res.},
  volume     = {15},
  number     = {1},
  pages      = {1593--1623},
  issn       = {1532-4435}
}

@article{gariazzo2022,
  title      = {Neutrino Mass and Mass Ordering: No Conclusive Evidence for Normal Ordering},
  shorttitle = {Neutrino Mass and Mass Ordering},
  author     = {Gariazzo, Stefano and Gerbino, Martina and Brinckmann, Thejs and Lattanzi, Massimiliano and Mena, Olga and Schwetz, Thomas and Choudhury, Shouvik Roy and Freese, Katherine and Hannestad, Steen and Ternes, Christoph A. and T{\'o}rtola, Mariam},
  year       = 2022,
  month      = oct,
  journal    = {Journal of Cosmology and Astroparticle Physics},
  volume     = {2022},
  number     = {10},
  pages      = {010},
  publisher  = {IOP Publishing},
  issn       = {1475-7516},
  doi        = {10.1088/1475-7516/2022/10/010},
  urldate    = {2025-11-26},
  langid     = {english}
}

@article{jimenez2022,
  title         = {Neutrino {{Masses}} and {{Mass Hierarchy}}: {{Evidence}} for the {{Normal Hierarchy}}},
  shorttitle    = {Neutrino {{Masses}} and {{Mass Hierarchy}}},
  author        = {Jimenez, Raul and {Pena-Garay}, Carlos and Short, Kathleen and Simpson, Fergus and Verde, Licia},
  year          = 2022,
  month         = sep,
  journal       = {Journal of Cosmology and Astroparticle Physics},
  volume        = {2022},
  number        = {09},
  eprint        = {2203.14247},
  primaryclass  = {hep-ph},
  pages         = {006},
  issn          = {1475-7516},
  doi           = {10.1088/1475-7516/2022/09/006},
  urldate       = {2025-10-01},
  archiveprefix = {arXiv}
}

@misc{schwetz2017,
  title         = {Comment on "{{Strong Evidence}} for the {{Normal Neutrino Hierarchy}}"},
  author        = {Schwetz, T. and Freese, K. and Gerbino, M. and Giusarma, E. and Hannestad, S. and Lattanzi, M. and Mena, O. and Vagnozzi, S.},
  year          = 2017,
  month         = mar,
  number        = {arXiv:1703.04585},
  eprint        = {1703.04585},
  primaryclass  = {astro-ph},
  publisher     = {arXiv},
  doi           = {10.48550/arXiv.1703.04585},
  urldate       = {2025-10-06},
  archiveprefix = {arXiv}
}

@article{simpson2017,
  title         = {Strong {{Bayesian Evidence}} for the {{Normal Neutrino Hierarchy}}},
  author        = {Simpson, Fergus and Jimenez, Raul and {Pena-Garay}, Carlos and Verde, Licia},
  year          = 2017,
  month         = jun,
  journal       = {Journal of Cosmology and Astroparticle Physics},
  volume        = {2017},
  number        = {06},
  eprint        = {1703.03425},
  primaryclass  = {astro-ph},
  pages         = {029--029},
  issn          = {1475-7516},
  doi           = {10.1088/1475-7516/2017/06/029},
  urldate       = {2025-10-01},
  archiveprefix = {arXiv}
}

@article{llorente2023,
  title         = {Marginal Likelihood Computation for Model Selection and Hypothesis Testing: An Extensive Review},
  shorttitle    = {Marginal Likelihood Computation for Model Selection and Hypothesis Testing},
  author        = {Llorente, Fernando and Martino, Luca and Delgado, David and {Lopez-Santiago}, Javier},
  year          = 2023,
  month         = feb,
  journal       = {SIAM Review},
  volume        = {65},
  number        = {1},
  eprint        = {2005.08334},
  primaryclass  = {stat},
  pages         = {3--58},
  issn          = {0036-1445, 1095-7200},
  doi           = {10.1137/20M1310849},
  urldate       = {2025-11-17},
  archiveprefix = {arXiv}
}

@article{piironen2017,
  title    = {Comparison of {{Bayesian}} Predictive Methods for Model Selection},
  author   = {Piironen, Juho and Vehtari, Aki},
  year     = 2017,
  month    = may,
  journal  = {Statistics and Computing},
  volume   = {27},
  number   = {3},
  pages    = {711--735},
  issn     = {1573-1375},
  doi      = {10.1007/s11222-016-9649-y},
  urldate  = {2025-11-27},
  langid   = {english}
}

@article{gelman2014,
  title    = {Understanding Predictive Information Criteria for {{Bayesian}} Models},
  author   = {Gelman, Andrew and Hwang, Jessica and Vehtari, Aki},
  year     = 2014,
  month    = nov,
  journal  = {Statistics and Computing},
  volume   = {24},
  number   = {6},
  pages    = {997--1016},
  issn     = {1573-1375},
  doi      = {10.1007/s11222-013-9416-2},
  urldate  = {2025-11-27},
  langid   = {english}
}

@article{sinharay2002,
  title      = {On the {{Sensitivity}} of {{Bayes Factors}} to the {{Prior Distributions}}},
  author     = {Sinharay, Sandip and Stern, Hal S.},
  year       = 2002,
  journal    = {The American Statistician},
  volume     = {56},
  number     = {3},
  eprint     = {3087298},
  eprinttype = {jstor},
  pages      = {196--201},
  publisher  = {[American Statistical Association, Taylor \& Francis, Ltd.]},
  issn       = {0003-1305},
  urldate    = {2025-11-27}
}

@incollection{gustafson2000,
  title     = {Local Robustness in Bayesian Analysis},
  booktitle = {Robust Bayesian Analysis},
  author    = {Gustafson, Paul},
  editor    = {Insua, David R{\'i}os and Ruggeri, Fabrizio},
  year      = 2000,
  pages     = {71--88},
  publisher = {Springer New York},
  address   = {New York, NY},
  doi       = {10.1007/978-1-4612-1306-2_4},
  isbn      = {978-1-4612-1306-2}
}

@article{kallioinen2023,
  title      = {Detecting and Diagnosing Prior and Likelihood Sensitivity with Power-Scaling},
  shorttitle = {Priorsense ({{SIR}})},
  author     = {Kallioinen, Noa and Paananen, Topi and B{\"u}rkner, Paul-Christian and Vehtari, Aki},
  year       = 2023,
  month      = dec,
  journal    = {Statistics and Computing},
  volume     = {34},
  number     = {1},
  pages      = {57},
  issn       = {1573-1375},
  doi        = {10.1007/s11222-023-10366-5},
  urldate    = {2025-10-01},
  langid     = {english}
}

@article{ohagan1995,
  title   = {Fractional Bayes Factors for Model Comparison},
  author  = {O'Hagan, Anthony},
  year    = 1995,
  journal = {Journal of the Royal Statistical Society: Series B (Methodological)},
  volume  = {57},
  number  = {1},
  eprint  = {https://academic.oup.com/jrsssb/article-pdf/57/1/99/49173542/jrsssb\_57\_1\_99.pdf},
  pages   = {99--118},
  issn    = {0035-9246},
  doi     = {10.1111/j.2517-6161.1995.tb02017.x}
}

@article{berger1996,
  title      = {The {{Intrinsic Bayes Factor}} for {{Model Selection}} and {{Prediction}}},
  author     = {Berger, James O. and Pericchi, Luis R.},
  year       = 1996,
  journal    = {Journal of the American Statistical Association},
  volume     = {91},
  number     = {433},
  eprint     = {2291387},
  eprinttype = {jstor},
  pages      = {109--122},
  publisher  = {[American Statistical Association, Taylor \& Francis, Ltd.]},
  issn       = {0162-1459},
  doi        = {10.2307/2291387},
  urldate    = {2025-11-28}
}

@article{gelfand1994,
  title      = {Bayesian {{Model Choice}}: {{Asymptotics}} and {{Exact Calculations}}},
  shorttitle = {Bayesian {{Model Choice}}},
  author     = {Gelfand, A. E. and Dey, D. K.},
  year       = 1994,
  month      = jan,
  journal    = {Journal of the Royal Statistical Society: Series B (Methodological)},
  volume     = {56},
  number     = {3},
  pages      = {501--514},
  issn       = {0035-9246},
  doi        = {10.1111/j.2517-6161.1994.tb01996.x},
  urldate    = {2025-11-28}
}

@article{metodiev2025,
  title     = {Easily {{Computed Marginal Likelihoods}} from {{Posterior Simulation Using}} the {{THAMES Estimator}}},
  author    = {Metodiev, Martin and {Perrot-Dock{\`e}s}, Marie and Ouadah, Sarah and Irons, Nicholas J. and Latouche, Pierre and Raftery, Adrian E.},
  year      = 2025,
  month     = sep,
  journal   = {Bayesian Analysis},
  volume    = {20},
  number    = {3},
  pages     = {1003--1030},
  publisher = {International Society for Bayesian Analysis},
  issn      = {1936-0975, 1931-6690},
  doi       = {10.1214/24-BA1422},
  urldate   = {2025-09-02}
}

@article{robert2009,
  title   = {Computational Methods for {{Bayesian}} Model Choice},
  author  = {Robert, C. P. and Wraith, D.},
  year    = 2009,
  month   = dec,
  journal = {AIP Conference Proceedings},
  volume  = {1193},
  number  = {1},
  pages   = {251--262},
  issn    = {0094-243X},
  doi     = {10.1063/1.3275622},
  urldate = {2025-11-28}
}

@inproceedings{polanska2024,
  author    = {A.~Polanska and T.~Wouters and T. H. Pang and
               K.~W.~K.~Wong and J.~D.~McEwen},
  booktitle = {Proceedings of the Machine Learning and Physical Sciences
               Workshop as part of the 38th International Conference on Neural
               Information Processing Systems (NeurIPS)},
  eprint    = {arXiv:2410.21076},
  location  = {Vancouver, Canada},
  month     = {Sep},
  projects  = {harmonic},
  title     = {Accelerated Bayesian parameter estimation and model selection for
               gravitational waves with normalizing flows},
  year      = {2024}
}

@article{piras2024,
  title         = {The Future of Cosmological Likelihood-Based Inference: Accelerated High-Dimensional Parameter Estimation and Model Comparison},
  shorttitle    = {The Future of Cosmological Likelihood-Based Inference},
  author        = {Piras, Davide and Polanska, Alicja and Mancini, Alessio Spurio and Price, Matthew A. and McEwen, Jason D.},
  year          = 2024,
  month         = sep,
  journal       = {The Open Journal of Astrophysics},
  volume        = {7},
  eprint        = {2405.12965},
  primaryclass  = {astro-ph},
  issn          = {2565-6120},
  doi           = {10.33232/001c.123368},
  urldate       = {2025-05-14},
  archiveprefix = {arXiv}
}

@article{spuriomancini2023,
  title   = {Bayesian Model Comparison for Simulation-Based Inference},
  author  = {Spurio~Mancini, A and Docherty, M M and Price, M A and McEwen, J D},
  year    = 2023,
  month   = jan,
  journal = {RAS Techniques and Instruments},
  volume  = {2},
  number  = {1},
  pages   = {710--722},
  issn    = {2752-8200},
  doi     = {10.1093/rasti/rzad051},
  urldate = {2024-10-09}
}

@article{gronau2020,
  title      = {Bridgesampling: {{An R Package}} for {{Estimating Normalizing Constants}}},
  shorttitle = {Bridgesampling},
  author     = {Gronau, Quentin F. and Singmann, Henrik and Wagenmakers, Eric-Jan},
  year       = 2020,
  month      = feb,
  journal    = {Journal of Statistical Software},
  volume     = {92},
  pages      = {1--29},
  issn       = {1548-7660},
  doi        = {10.18637/jss.v092.i10},
  urldate    = {2025-11-28},
  copyright  = {Copyright (c) 2020 Quentin F. Gronau, Henrik Singmann, Eric-Jan Wagenmakers},
  langid     = {english}
}

@article{berger1994,
  title    = {An Overview of Robust {{Bayesian}} Analysis},
  author   = {Berger, James O. and Moreno, El{\'i}as and Pericchi, Luis Raul and Bayarri, M. Jes{\'u}s and Bernardo, Jos{\'e} M. and Cano, Juan A. and {De la Horra}, Juli{\'a}n and Mart{\'i}n, Jacinto and {R{\'i}os-Ins{\'u}a}, David and Betr{\`o}, Bruno and Dasgupta, A. and Gustafson, Paul and Wasserman, Larry and Kadane, Joseph B. and Srinivasan, Cid and Lavine, Michael and O'Hagan, Anthony and Polasek, Wolfgang and Robert, Christian P. and Goutis, Constantinos and Ruggeri, Fabrizio and Salinetti, Gabriella and Sivaganesan, Siva},
  year     = 1994,
  month    = jun,
  journal  = {Test},
  volume   = {3},
  number   = {1},
  pages    = {5--124},
  issn     = {1863-8260},
  doi      = {10.1007/BF02562676},
  urldate  = {2025-11-28},
  langid   = {english}
}

@unpublished{hu2026,
  author = {Hu, Zixiao and McEwen, Jason D.},
  note   = {In preparation},
  title  = {Fast and flexible {Bayesian} model comparison with emulated likelihoods},
  year   = {2026}
}

@misc{naderi2025,
  title         = {Approximating Evidence via Bounded Harmonic Means},
  author        = {Naderi, Dana and Robert, Christian P. and Kamary, Kaniav and Wraith{\S}, Darren},
  year          = 2025,
  month         = oct,
  number        = {arXiv:2510.20617},
  eprint        = {2510.20617},
  primaryclass  = {stat},
  publisher     = {arXiv},
  doi           = {10.48550/arXiv.2510.20617},
  urldate       = {2025-12-09},
  archiveprefix = {arXiv}
}

@article{carrion2025,
  title   = {Testing Interacting Dark Energy with {{Stage IV}} Cosmic Shear Surveys through Differentiable Neural Emulators},
  author  = {Carrion, Karim and Spurio~Mancini, Alessio and Piras, Davide and Hidalgo, Juan Carlos},
  year    = 2025,
  month   = jun,
  journal = {Monthly Notices of the Royal Astronomical Society},
  volume  = {539},
  number  = {4},
  pages   = {3220--3228},
  issn    = {0035-8711},
  doi     = {10.1093/mnras/staf663},
  urldate = {2025-12-09}
}

@article{paradiso2024,
  title      = {Evaluating Extensions to {{LCDM}}: An Application of {{Bayesian}} Model Averaging and Selection},
  shorttitle = {Evaluating Extensions to {{LCDM}}},
  author     = {Paradiso, S. and McGee, G. and Percival, W.J.},
  year       = 2024,
  month      = oct,
  journal    = {Journal of Cosmology and Astroparticle Physics},
  volume     = {2024},
  number     = {10},
  pages      = {021},
  publisher  = {IOP Publishing},
  issn       = {1475-7516},
  doi        = {10.1088/1475-7516/2024/10/021},
  urldate    = {2025-12-09},
  langid     = {english}
}

@article{smith1992,
  title      = {Bayesian {{Statistics}} without {{Tears}}: {{A Sampling-Resampling Perspective}}},
  shorttitle = {Bayesian {{Statistics}} without {{Tears}}},
  author     = {Smith, A. F. M. and Gelfand, A. E.},
  year       = 1992,
  journal    = {The American Statistician},
  volume     = {46},
  number     = {2},
  eprint     = {2684170},
  eprinttype = {jstor},
  pages      = {84--88},
  publisher  = {[American Statistical Association, Taylor \& Francis, Ltd.]},
  issn       = {0003-1305},
  doi        = {10.2307/2684170},
  urldate    = {2024-11-01},
  lccn       = {1539}
}

@article{rubin1987,
  title      = {The {{Calculation}} of {{Posterior Distributions}} by {{Data Augmentation}}: {{Comment}}: {{A Noniterative Sampling}}/{{Importance Resampling Alternative}} to the {{Data Augmentation Algorithm}} for {{Creating}} a {{Few Imputations When Fractions}} of {{Missing Information Are Modest}}: {{The SIR Algorithm}}},
  shorttitle = {The {{Calculation}} of {{Posterior Distributions}} by {{Data Augmentation}}},
  author     = {Rubin, Donald B.},
  year       = 1987,
  journal    = {Journal of the American Statistical Association},
  volume     = {82},
  number     = {398},
  eprint     = {2289460},
  eprinttype = {jstor},
  pages      = {543--546},
  publisher  = {[American Statistical Association, Taylor \& Francis, Ltd.]},
  issn       = {0162-1459},
  doi        = {10.2307/2289460},
  urldate    = {2024-10-16},
  lccn       = {154}
}

@article{stiskalek2025,
  title      = {The {{Velocity Field Olympics}}: {{Assessing}} Velocity Field Reconstructions with Direct Distance Tracers},
  shorttitle = {The {{Velocity Field Olympics}}},
  author     = {Stiskalek, Richard and Desmond, Harry and Devriendt, Julien and Slyz, Adrianne and Lavaux, Guilhem and Hudson, Michael J. and Bartlett, Deaglan J. and Courtois, H{\'e}l{\`e}ne M.},
  year       = 2025,
  month      = nov,
  journal    = {Monthly Notices of the Royal Astronomical Society},
  publisher  = {OUP},
  issn       = {0035-8711},
  doi        = {10.1093/mnras/staf1960},
  urldate    = {2025-12-10}
}

@article{stiskalek2024,
  title     = {Symmetry in {{Hyper Suprime-Cam Galaxy Spin Directions}}},
  author    = {Stiskalek, Richard and Desmond, Harry},
  year      = 2024,
  month     = nov,
  journal   = {Research Notes of the AAS},
  volume    = {8},
  number    = {11},
  pages     = {281},
  publisher = {The American Astronomical Society},
  issn      = {2515-5172},
  doi       = {10.3847/2515-5172/ad8fb1},
  urldate   = {2025-12-10},
  langid    = {english}
}

@article{du2025,
  title   = {Impacts of Dark Energy on Weighing Neutrinos after {{DESI BAO}}},
  author  = {Du, Guo-Hong and Wu, Peng-Ju and Li, Tian-Nuo and Zhang, Xin},
  year    = 2025,
  month   = apr,
  journal = {The European Physical Journal C},
  volume  = {85},
  number  = {4},
  pages   = {392},
  issn    = {1434-6052},
  doi     = {10.1140/epjc/s10052-025-14094-0},
  urldate = {2025-12-10},
  langid  = {english}
}

@article{papamakarios2021,
  title   = {Normalizing {{Flows}} for {{Probabilistic Modeling}} and {{Inference}}},
  author  = {Papamakarios, George and Nalisnick, Eric and Rezende, Danilo Jimenez and Mohamed, Shakir and Lakshminarayanan, Balaji},
  year    = 2021,
  journal = {Journal of Machine Learning Research},
  volume  = {22},
  number  = {57},
  pages   = {1--64},
  issn    = {1533-7928},
  urldate = {2025-12-10}
}

@misc{spuriomancini2024,
  title      = {Field-Level Cosmological Model Selection: Field-Level Simulation-Based Inference for {{Stage IV}} Cosmic Shear Can Distinguish Dynamical Dark Energy},
  shorttitle = {Field-Level Cosmological Model Selection},
  author     = {Spurio Mancini, A. and Lin, K. and McEwen, J. D.},
  year       = 2024,
  month      = oct,
  publisher  = {arXiv},
  doi        = {10.48550/arXiv.2410.10616},
  urldate    = {2025-12-10}
}

@misc{rinaldi2024,
  title         = {Hierarchical Inference of Evidence Using Posterior Samples},
  author        = {Rinaldi, Stefano and Demasi, Gabriele and Pozzo, Walter Del and Hannuksela, Otto A.},
  year          = 2024,
  month         = may,
  number        = {arXiv:2405.07504},
  eprint        = {2405.07504},
  primaryclass  = {stat},
  publisher     = {arXiv},
  doi           = {10.48550/arXiv.2405.07504},
  urldate       = {2025-12-10},
  archiveprefix = {arXiv}
}

@article{srinivasan2024,
  title         = {Bayesian Evidence Estimation from Posterior Samples with Normalizing Flows},
  shorttitle    = {Density Estimation},
  author        = {Srinivasan, Rahul and Crisostomi, Marco and Trotta, Roberto and Barausse, Enrico and Breschi, Matteo},
  year          = 2024,
  month         = dec,
  journal       = {Physical Review D},
  volume        = {110},
  number        = {12},
  eprint        = {2404.12294},
  primaryclass  = {stat},
  pages         = {123007},
  issn          = {2470-0010, 2470-0029},
  doi           = {10.1103/PhysRevD.110.123007},
  urldate       = {2025-09-16},
  archiveprefix = {arXiv}
}

@misc{heavens2017,
  title         = {Marginal {{Likelihoods}} from {{Monte Carlo Markov Chains}}},
  author        = {Heavens, Alan and Fantaye, Yabebal and Mootoovaloo, Arrykrishna and Eggers, Hans and Hosenie, Zafiirah and Kroon, Steve and Sellentin, Elena},
  year          = 2017,
  month         = apr,
  number        = {arXiv:1704.03472},
  eprint        = {1704.03472},
  primaryclass  = {stat},
  publisher     = {arXiv},
  doi           = {10.48550/arXiv.1704.03472},
  urldate       = {2025-12-10},
  archiveprefix = {arXiv}
}

@inproceedings{jia2020,
  title     = {Normalizing {{Constant Estimation}} with {{Gaussianized Bridge Sampling}}},
  booktitle = {Proceedings of {{The}} 2nd {{Symposium}} on  {{Advances}} in {{Approximate Bayesian Inference}}},
  author    = {Jia, He and Seljak, Uros},
  year      = 2020,
  month     = feb,
  pages     = {1--14},
  publisher = {PMLR},
  issn      = {2640-3498},
  urldate   = {2025-12-10},
  langid    = {english}
}

@article{carrillo2014purify,
  title     = {PURIFY: a new approach to radio-interferometric imaging},
  author    = {Carrillo, Rafael E and McEwen, Jason D and Wiaux, Yves},
  journal   = {Monthly Notices of the Royal Astronomical Society},
  volume    = {439},
  number    = {4},
  pages     = {3591--3604},
  year      = {2014},
  publisher = {Oxford Academic}
}

@book{lee1989bayesian,
  title     = {Bayesian statistics},
  author    = {Lee, Peter M},
  year      = {1989},
  publisher = {Oxford University Press London:}
}

@article{price2021sparse,
  title     = {Sparse Bayesian mass-mapping with uncertainties: Full sky
               observations on the celestial sphere},
  author    = {Price, Matthew A and McEwen, Jason D and Pratley, L and Kitching,
               Thomas D},
  journal   = {Monthly Notices of the Royal Astronomical Society},
  volume    = {500},
  number    = {4},
  pages     = {5436--5452},
  year      = {2021},
  publisher = {Oxford University Press}
}

@article{remy2023probabilistic,
  title     = {Probabilistic mass-mapping with neural score estimation},
  author    = {Remy, Benjamin and Lanusse, Francois and Jeffrey, Niall and Liu, Jia
               and Starck, J-L and Osato, Ken and Schrabback, Tim},
  journal   = {Astronomy \& Astrophysics},
  volume    = {672},
  pages     = {A51},
  year      = {2023},
  publisher = {EDP Sciences}
}

@article{alsing2021nested,
  title     = {Nested sampling with any prior you like},
  author    = {Alsing, Justin and Handley, Will},
  journal   = {Monthly Notices of the Royal Astronomical Society: Letters},
  volume    = {505},
  number    = {1},
  pages     = {L95--L99},
  year      = {2021},
  publisher = {Oxford University Press}
}

@article{liaudat2024scalable,
  title     = {Scalable Bayesian uncertainty quantification with data-driven priors
               for radio interferometric imaging},
  author    = {Liaudat, Tob{\'\i}as I and Mars, Matthijs and Price, Matthew A and
               Pereyra, Marcelo and Betcke, Marta M and McEwen, Jason D},
  journal   = {RAS Techniques and Instruments},
  volume    = {3},
  number    = {1},
  pages     = {505--534},
  year      = {2024},
  publisher = {Oxford University Press}
}

@inproceedings{mcewen2023proximal,
  title        = {Proximal nested sampling with data-driven priors for physical
                  scientists},
  author       = {McEwen, Jason D and Liaudat, Tob{\'\i}as I and Price, Matthew A and
                  Cai, Xiaohao and Pereyra, Marcelo},
  booktitle    = {Physical Sciences Forum},
  volume       = {9},
  number       = {1},
  pages        = {13},
  year         = {2023},
  organization = {MDPI}
}

@inproceedings{skilling2004,
  title      = {Nested {{Sampling}}},
  booktitle  = {Bayesian {{Inference}} and {{Maximum Entropy Methods}} in {{Science}} and {{Engineering}}: 24th {{International Workshop}} on {{Bayesian Inference}} and {{Maximum Entropy Methods}} in {{Science}} and {{Engineering}}},
  author     = {Skilling, John},
  year       = 2004,
  month      = nov,
  volume     = {735},
  pages      = {395--405},
  publisher  = {AIP},
  doi        = {10.1063/1.1835238},
  urldate    = {2025-12-11}
}

@article{meng1996,
  title      = {Simulating {{Ratios}} of {{Normalizing Constants Via}} a {{Simple Identity}}: {{A Theoretical Exploration}}},
  shorttitle = {Simulating {{Ratios}} of {{Normalizing Constants Via}} a {{Simple Identity}}},
  author     = {Meng, Xiao-Li and Wong, Wing Hung},
  year       = 1996,
  journal    = {Statistica Sinica},
  volume     = {6},
  number     = {4},
  eprint     = {24306045},
  eprinttype = {jstor},
  pages      = {831--860},
  publisher  = {Institute of Statistical Science, Academia Sinica},
  issn       = {1017-0405},
  urldate    = {2025-12-11}
}

@article{whitney2025,
  title   = {Generative Modelling for Mass-Mapping with Fast Uncertainty Quantification},
  author  = {Whitney, Jessica J and Liaudat, Tob{\'i}as I and Price, Matthew A and Mars, Matthijs and McEwen, Jason D},
  year    = 2025,
  month   = sep,
  journal = {Monthly Notices of the Royal Astronomical Society},
  volume  = {542},
  number  = {3},
  pages   = {2464--2479},
  issn    = {0035-8711},
  doi     = {10.1093/mnras/staf1356},
  urldate = {2025-12-11}
}

\end{document}